\documentclass[ preprintnumbers,amsmath]{revtex4}
\usepackage{epsfig}
\usepackage{bm}
\usepackage{amsmath}
\usepackage{amssymb}
\usepackage{epstopdf}
\begin{document}
\title{Comment on ''Partial energies fluctuations and negative heat capacities'' by
X. Campi et al.}
\author{Ph. Chomaz}
\affiliation{GANIL ( DSM - CEA / IN2P3 - CNRS),
B.P.5027, F-14021 Caen C\'edex, France} 
\author{F.\ Gulminelli} 
\altaffiliation{member of the Institut Universitaire de France}
\affiliation{
LPC Caen (IN2P3 - CNRS / EnsiCaen et Universit\'{e}), F-14050
Caen C\'{e}dex, France} 
\author{M.D'Agostino} 
\affiliation{Dipartimento di Fisica and INFN, Bologna, Italy}

\begin{abstract}
Studying the energy partioning published by \cite{Campi2004} we show that
the presented results do not fulfill the 
sum rule due to energy
conservation. The 
observed fluctuations of the energy conservation test 
point to a 
numerical problem.
Moreover, analysis of the binding energies show that the 
fragment recognition algorithm adopted 
by Campi et al. leads with a sizeable probability 
to fragments containing up to the total mass even for excitation energies
as large as 3/4 of the total binding. This surprising result points to 
another problem since the published inter-fragment energy is not zero 
while a unique fragment is present. This problem may be due to 
either the 
fragment recognition algorithm or to the definition of the inter and 
intra-fragment energy. These numerical inconsistencies 
should be settled before any conclusion on the physics can be drawn. 
\end{abstract}

\maketitle
 
In a recent article\cite{Campi2004} X. Campi et al. present
results from numerical experiments on the liquid gas phase 
transition of a Van der Vaals fluid using a molecular dynamics simulations. 
They acknowledge the fact that in a microcanonical ensemble 
the kinetic energy fluctuation 
($\sigma _{K}^{2}/T^{2}$) 
can be used to reconstruct the heat capacities in 
the context of phase transitions as proposed in ref.\cite{NegC-theo}, but
they criticize the experimental evaluation of such a quantity from the
observed partitions as used in ref. \cite{NegC-exp}.

Before discussing the actual criticisms of Campi et al. to our work,
we would like to point out that the numerical results presented in ref.  
\cite{Campi2004} are inconsistent.

Campi et al. present 4 quantities, 

\begin{itemize}
\item  the exact kinetic energy $K=E-V$,
where $E$ is the total energy and $V$ the total interaction energy, 
\item  the quantity supposed to be the approximation to the kinetic energy
introduced in ref. \cite{NegC-exp}, 
$K^{\prime }=E-\sum_{i}B_{i}$
where the sum runs over the fragments and where $B_{i}$ are the fragment
ground state energy \cite{hill}, 
\item  the inter-fragment interaction 
$V_{inter}=\sum_{i<j}V_{ij}$
\item  and the intra-fragment energy 
$V_{intra}=\sum_{i}V_{i}$
which they show as a configurational excitation energy
$\Delta V=\sum_{i}V_{i}-B_{i}$
\end{itemize}

Since $V=V_{inter}+$ $V_{intra}$ the above quantities should at any time
fulfill the exact sum rule 
\begin{equation}
S=K-K^{\prime }+V_{inter}+\Delta V=0  \label{Sum-rule}
\end{equation}
The right side of figure 1 shows the sum rule (\ref{Sum-rule}) as obtained
from the data of ref. \cite{Campi2004}, which are 
shown in the left part of the same figure.
\begin{figure}[tbh]
\begin{center}
\includegraphics[width=8cm]{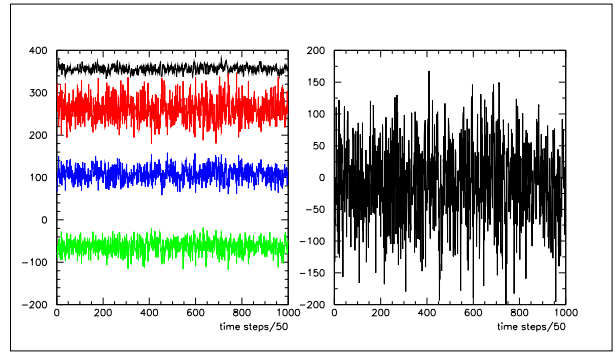}
\end{center}
\caption{Left side: Fig 3 of \protect\cite{Campi2004} 
showing from bottom to top 
$V_{inter}$ , $\Delta V$ , $K^{\prime }$ shifted by $100$ units , 
$K$ shifted by $250$ units. Right side: the corresponding energy sum 
rule $S=K-K^{\prime}+V_{inter}+\Delta V$.}
\label{fig:1}
\end{figure}

Fig 1 shows that in the Campi et al. calculations the sum rule $S$ 
is not zero, suggesting a possible numerical mistake. 
If the error on the average $<S>=-8.871$ is of the order of a few percent,
the more worrying, considering that the aim of the Campi et al. paper is to
discuss fluctuations, are the enormous fluctuations presented by this quantity 
which should be exactly conserved in time. 
Indeed the observed variance of S is $3819$
while the one of $K^{\prime }$ criticized by Campi et al. to be too large is
only $784.$ 
The problem can be more easily understood looking at figure 2 which 
is the same as figure 3 of ref.\cite{Campi2004} on an enlarged time scale.
\begin{figure}[tbh]
\begin{center}
\includegraphics[width=8cm]{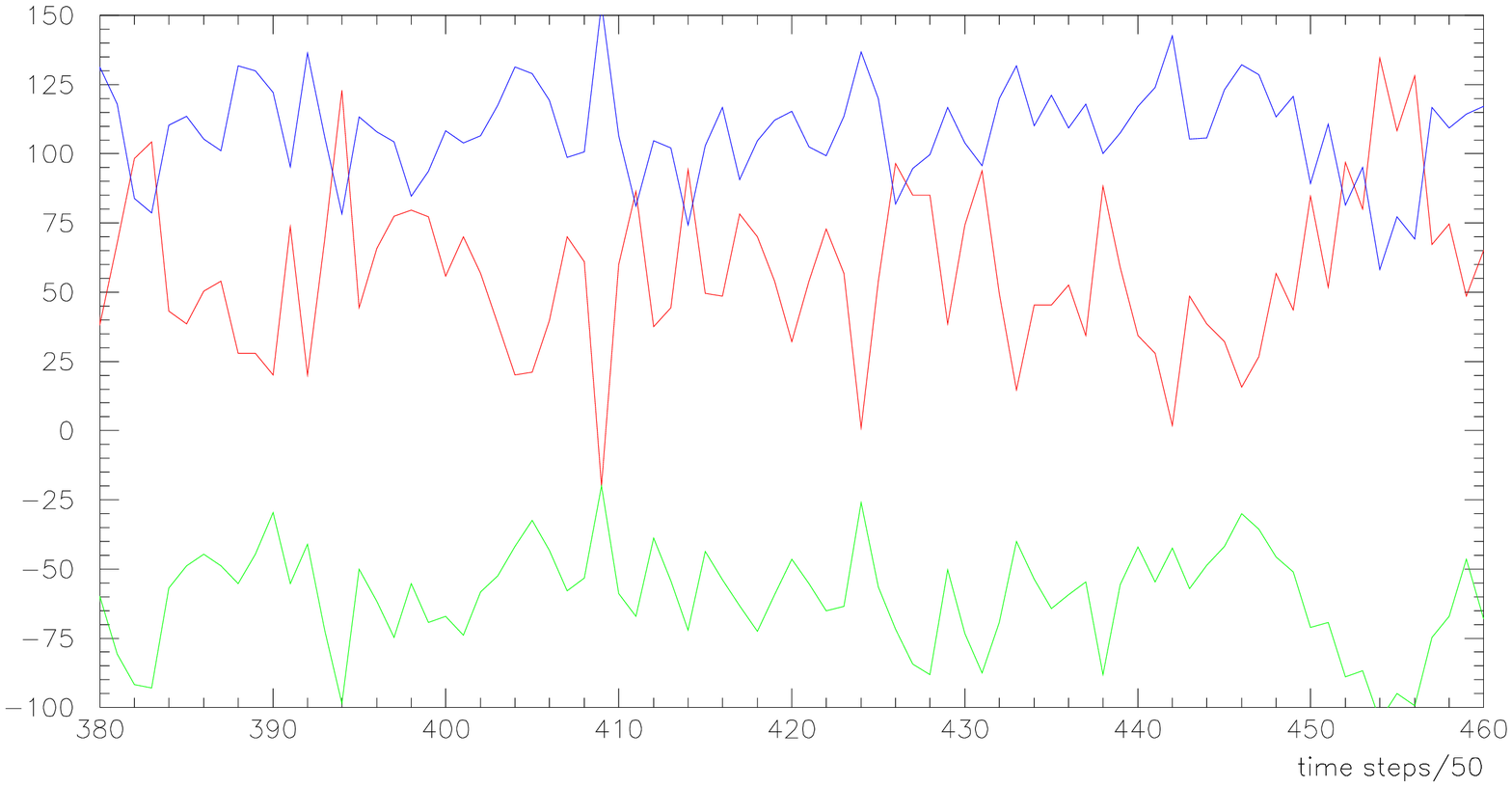}
\end{center}
\caption{Same as the left part of Figure 1 on an enlarged scale,
showing from bottom to top $V_{inter}$,
$K^{\prime }$ shifted by $-100$ units and $\Delta V$.}
\label{fig:2}
\end{figure}
In this figure we can see that the fluctuations of $\Delta V$ and $V_{inter}$
are approximately in phase.
If we consider that the fluctuations of $K$ are very small as shown in figure 1,
for the sum rule to be exactly fulfilled, $K^{\prime }$ should be 
positively correlated with $\Delta V$ and $V_{inter}$ 
which is clearly not the case in the calculation of Campi et al.

Therefore, before entering into any discussions about the
physical meaning of Campi et al. findings, one should be sure that the 
numerics is under control.

\begin{figure}[tbh]
\begin{center}
\includegraphics[width=8cm]{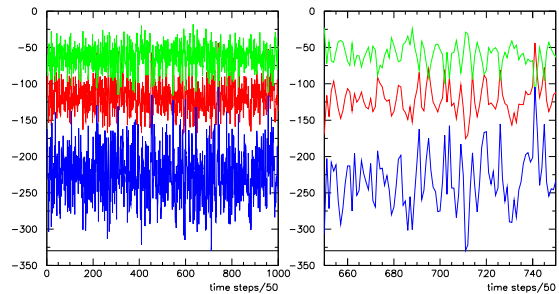}
\end{center}
\caption{Left side: from bottom to top, the
ground state binding energy of the N=64 system $B_{64}$ from
ref.\protect\cite{hill}, the binding
energy of the partitions $B,$ the intra-fragment energy $V_{intra}$ and the
inter-fragment energy $V_{inter}$, for the same data as in figure 1.
Right side: same as the left side, but on an enlarged time scale. }
\label{fig:3e}
\end{figure}
In order to understand better what might be going on, we have used $K,$ 
$V_{inter}$ and $\Delta V$ combined with the Lennard Jones mass 
table\cite{hill} and the excitation energy reported in \cite{Campi2004} 
to disentangle the various components entering in the calculation 
and in particular the total binding energy of the partitioned system, 
$B=\sum_i B_i$. 
We have thus used for the total energy $E=E^{*}$ $+$ $B_{64}=-73.62$ 
since $E^{*}/N=4$ according to 
\cite{Campi2004} and since $B_{64}=-329.62$ according to the  mass 
table\cite{hill}. Then we can get $V$ from $K=E-V$ and thus $V_{intra}$ from 
$V=V_{inter}+$ $V_{intra}$ and finally $B$ from $\Delta V=V_{intra}-B.$ 
The results are shown in figure 3. Before discussing the fact that $B$ 
is a very poor approximation of $V_{intra}$ in the calculations performed 
by Campi et al., one is surprised by the fact that $B$ is reaching exactly 
$B_{64}=-329.62$ at time step $711$ in the figure 3 of ref. \cite{Campi2004}. 
This means that there is a non negligible probability 
that the system is composed by 
a unique fragment containing exactly all the particles  
(for comparison $B_{63}=-323.49$). 
For this peculiar single-fragment event 
Campi et al. report a non zero
inter-fragment energy ($V_{inter}=-29$), which again points
to a numerical problem. 
This might be
related to the sum rule problem discussed above but it is likely to be a
different one, since the first problem is related to $K^{\prime }$ while in
the second one the information about $K^{\prime }$ is not used.
One of the problems may lie in the fragment definition algorithm
employed by Campi et al. 
In this respect it may be worthwhile to notice that, even if the 
classical Lennard Jones molecular dynamics is in itself an exact
well controlled model, on the other hand 
the definition of physical clusters in these kind of models 
and their relationship with measured fragments 
is subject to many controversies\cite{claudio}.

In conclusion, our study of the results published by Campi et al. clearly
show two numerical problems:
\begin{itemize}
\item   the sum rule taking care of the energy conservation in the energy
partition is violated and presents fluctuations, 
\item  the analysis of the total binding energy of the partitioned system
shows that even at excitation energies above the critical point, the
fragment recognition algorithm used in ref \cite{Campi2004} predicts the
existence of residues of very large masses, containing up to the 64 particles
of the considered system, and pointing to a problem in the definition of the
inter-fragment energy since in the case of a unique fragment this latter 
is not zero. 
\end{itemize}
These numerical inconsistencies 
should be settled before any conclusion on the physics can be drawn. 

To progress on this issue, we have repeated the Campi et alanalysis\cite{Campi2004} of the independent fragment approximation for the evaluation of the interaction energy on another well controlled exact model.  We show in a forthcoming paper that this approximation is accurate provided that the required consistency tests \cite{palluto} are taken into account.


\end{document}